\documentclass[prd,twocolumn,a4paper,superscriptaddress,nofootinbib]{revtex4-2}

\usepackage{amsmath,graphicx,amssymb,xcolor}
\usepackage[colorlinks=true, allcolors=blue]{hyperref}
\usepackage{soul}

\setstcolor{red}

\begin{document}

\title{Electroweak Axion as the Fuzzy Dark Matter\\
----- A Proposal for the Mixed Fuzzy and Cold Dark Matters -----}

\author{Yu-Cheng Qiu}
\email{ethanqiu@sjtu.edu.cn}
\affiliation{Tsung-Dao Lee Institute and School of Physics and Astronomy, \\
Shanghai Jiao Tong University, 520 Shengrong Road, Shanghai, 201210, China}

\author{Tsutomu T. Yanagida}
\email{tsutomu.tyanagida@sjtu.edu.cn}
\affiliation{Tsung-Dao Lee Institute and School of Physics and Astronomy, \\
Shanghai Jiao Tong University, 520 Shengrong Road, Shanghai, 201210, China}
\affiliation{Kavli IPMU (WPI), The University of Tokyo, Kashiwa, Chiba 277-8583, Japan}

\date{\today}

\begin{abstract}
The electroweak axion is identified with the fuzzy dark matter of a mass $m\simeq 10^{-20}$--$10^{-19}\,{\rm eV}$. The model predicts two components of dark matter, one is ultralight and the other is WIMP-like. The Chern-Simons-type interaction between the fuzzy dark matter and photon and the $B+L$ breaking proton decays are predicted.
\end{abstract}

\maketitle

\section{Introduction}\label{sec:intro}

The presence of axion-like Nambu-Goldstone bosons is a quite natural prediction of string theories~\cite{Svrcek:2006yi}. The QCD axion is one of them which couples to QCD instantons, solving the CP problem in QCD. This success encourages us to consider another axion which couples to weak $SU(2)$ instantons. The weak $SU(2)$ instantons generate a very small mass for the axion. We call it as the electroweak (EW) axion. 

The EW axion was originally introduced to explain the observed extremely small cosmological constant by its potential energy density~\cite{Nomura:2000yk}. 
In this paper, we propose to identify the EW axion with the Fuzzy dark matter (DM)~\cite{Hu:2000ke} instead of explaining the cosmological constant.

\section{A Brief Review on the Original Electroweak Axion}\label{sec:EW_axion}

In this section we give a brief review on the original EW axion introduced to explain the observed very small cosmological constant by the axion potential energy density~\cite{Nomura:2000yk}.
The EW axion $A$ couples to the $SU(2)$ gauge fields as
\begin{equation}
\label{eq:AWW-coupling}
\mathcal{L}\supset\frac{g^2_2}{32\pi^2} \frac{A}{F_A} W_{\mu \nu}^i\widetilde{W}^{i\mu \nu}\;,
\end{equation}
where $W_{\mu \nu}^i$ (with $i=1,2,3$) is the weak $SU(2)$ gauge field strength tensor, $\widetilde{W}^{i\mu \nu}$ its dual tensor and $g_2$ is the weak $SU(2)$ gauge coupling constant and $F_A$ is the decay constant of the EW axion. 
Here, we explain why the EW axion can explain the observed cosmological constant by its potential energy~\footnote{It is also shown to be able to explain the recently observed cosmic birefringence~\cite{Choi:2021aze,Lin:2022niw}.}. 

\begin{figure}
\includegraphics[width=8cm]{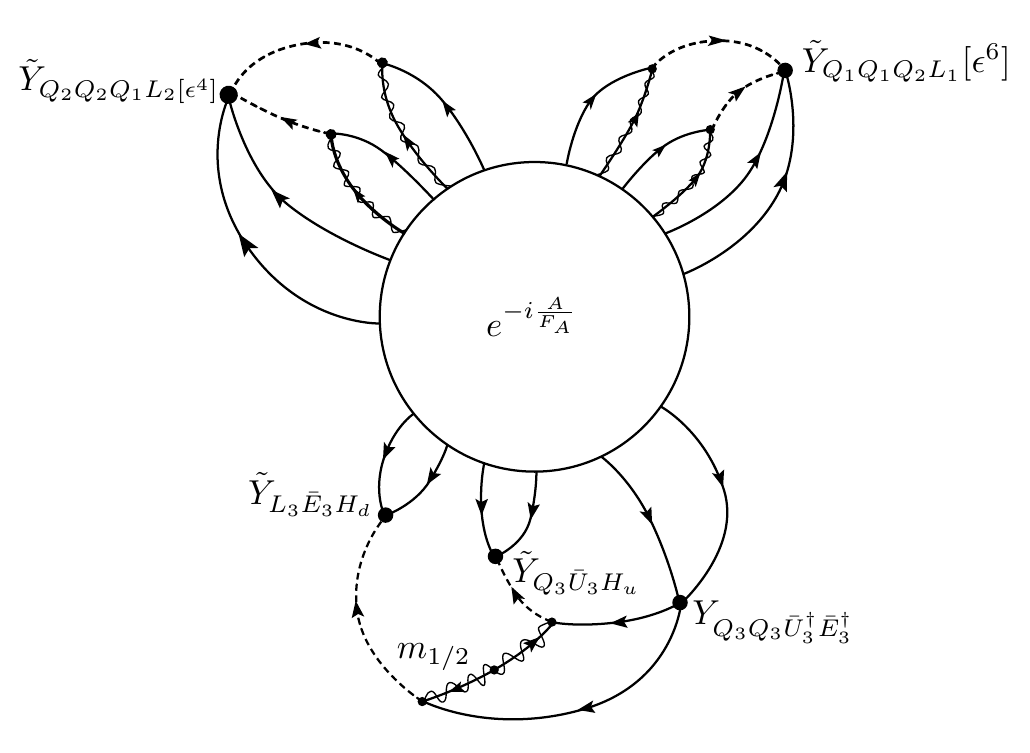}
\caption{One anti-instanton diagram generating the axion potential.}
\label{fig:ew_instanton}
\end{figure}

We have 18 fermion zero modes around an instanton as shown in Fig.~\ref{fig:ew_instanton} and we have to contract all fermion zero modes by higher dimensional operators to generate the axion potential~\cite{Nomura:2000yk}. In particular, the higher dimensional operators necessarily contain the $B+L$ breaking operators, which generate too fast proton decays~\cite{Sakai:1981pk, Weinberg:1981wj}. We need some flavor symmetry acting on the quarks and leptons to suppress such fast proton decays, but on the other hand, the flavor symmetry must be broken to generate the observed mass matrices of quarks and leptons. We adopt, in this section, the global Froggatt-Nielsen $U(1)_{\rm FN}$ flavor symmetry~\cite{Froggatt:1978nt}. We introduce a symmetry breaking spurion $\epsilon \simeq 1/17$~\cite{Buchmuller:1998zf} to generate the realistic mass matrices for quarks and leptons. 

With this spurion parameter $\epsilon$ the instanton calculus gives us the axion potential as~\cite{Nomura:2000yk}
\begin{equation}
\label{eq:axio-potential}
V_A=\frac{\Lambda_{A}^4}{2}\left[1-\cos\left(\frac{A}{F_A}\right)\right] \;,
\end{equation}
with
\begin{align}
\Lambda_{A}^4 & \simeq 2 e^{-\frac{2\pi}{\alpha_2(M_{\rm Pl})}} c\,\left(\frac{1}{\pi^2}\right)^4 \epsilon^{10} m_{3/2}^3 M_{\rm Pl} \label{eq:potential-height}\\
    &\simeq \left(1.4\times10^{-3}\,{\rm eV}\right)^4 c\left(\frac{1}{\pi^2}\right)^4 \left(\frac{\epsilon}{1/17}\right)^{10} \left(\frac{m_{3/2}}{1\,{\rm TeV}}\right)^3 \;, \nonumber
\end{align}
where $c$ is a dimensionless constant of $\mathcal{O}(1)$, $m_{3/2}$ is the gravitino mass, and $\alpha_2(M_{\rm Pl})$ is the weak $SU(2)$ gauge coupling constant at the Planck scale~\footnote{This result does not change even if some $SU(2)$ charged particles exist at the intermediate energy scale owing to the SUSY miracle~\cite{Nomura:2000yk}.}.  An example of the diagrams generating the above EW axion potential is also shown in Fig.1. One might think that the instanton contributions are very much suppressed by loop factors. However, the meaning of loops is not trivial in the instanton calculus. The correct $\pi^2$ factor is given in the appendix in \cite{Nomura:2000yk}, which gives us the factor $(\frac{1}{\pi^2})^4$ in Eq.~\eqref{eq:potential-height} in the present case. However, there are many contributions from all instanton diagrams and ambiguities coming from the effective coupling constants of the higher dimensional operators and the effective cutoff scale of the  instanton-size integrations. The constant $c$ represents such ambiguities. We take $c=\mathcal{O}(1)$ as a representative value throughout this paper.

The above EW axion potential Eq.~\eqref{eq:axio-potential} generates the correct observed dark energy at around the hilltop of the $\bf{cosin}$ potential in Eq.~\eqref{eq:axio-potential} for $c=\mathcal{O}(1)$ and $m_{3/2}=\mathcal{O}(10)\,{\rm TeV}$. Here, we have considered that the Planck scale $M_{\rm Pl} \simeq 2.4\times 10^{18}\,{\rm GeV}$ is the cut-off scale of the theory as explained in~\cite{Choi:2021aze}. 

From the axion potential Eq.~\eqref{eq:axio-potential} we obtain the EW axion mass $m_A$ around the potential minimum as
\begin{align}
m_A & =\frac{\Lambda_A^2}{\sqrt{2}F_A} \simeq 6\times10^{-34} \,{\rm eV} \nonumber\\
& \quad \times \left(\frac{1}{\pi^2}\right)^{2} \left(\frac{\epsilon}{1/17}\right)^{5} \left(\frac{m_{3/2}}{1\,{\rm TeV}}\right)^{3/2} \left(\frac{M_{\rm Pl}}{F_A}\right)  \;, \label{eq:axion-mass}
\end{align}
taking $c=1$.

\section{Electroweak Axion Potential With the Anomaly-Free Discrete Froggatt-Nielsen Symmetry}\label{sec:FN_symmetry}

\begin{figure}
\includegraphics[width=8cm]{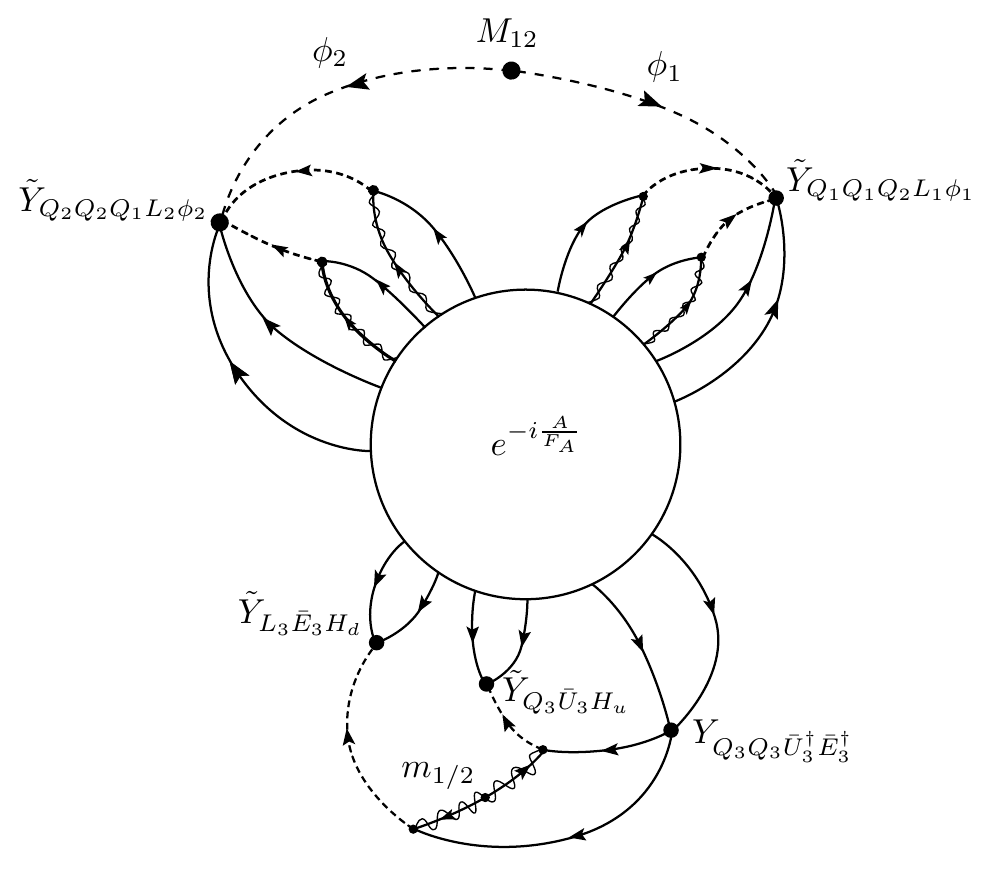}
\caption{One anti-instanton diagram with the discrete $Z_{10}$ FN symmetry and $M_{12}$-term introduced.}
\label{fig:ew_instanton_2}
\end{figure}

In the previous section, we have assumed the Froggatt-Nielsen $U(1)_{\rm FN}$ flavor symmetry. In this section, however, we consider the anomaly-free flavor symmetry, that is, the Froggatt-Nielsen discrete $Z_{10}$ symmetry~\cite{Choi:2019jck}.
There is a big difference in the axion potential from the estimation in the original paper since there is no suppression factor, $\epsilon^{10}$. Notice that the dangerous $B+L$ breaking operators for proton decays are still suppressed by powers of the $\epsilon$ as in the case of the $U(1)_{\rm FN}$. 
An example of the diagrams generating the EW axion potential is shown in Fig.~\ref{fig:ew_instanton_2}.

Now, we obtain the EW axion mass as
\begin{equation}
  m_A\simeq 1.2\times10^{-21} \sqrt{c} \left(\frac{1}{\pi^2}\right)^{5/2} \left(\frac{m_{3/2}}{1\,{\rm TeV}}\right)^{3/2}\left(\frac{M_{\rm Pl}}{F_A}\right) \,{\rm eV}\,.
 \end{equation}
Notice that the $\frac{1}{\pi^2}$ term becomes $(\frac{1}{\pi^2})^5$ in the axion potential because of the exchange of a new massive particle between two vertexes in Fig.~\ref{fig:ew_instanton_2}. Here, we introduce a pair of new singlet chiral multiplets, $\phi_1$ and $\phi_2$, which carry the Froggatt-Nielsen charges $-6$ and $-4$, respectively. (See Ref.~\cite{Nomura:2000yk} for the Froggatt-Nielsen charges of quarks and leptons.) And we consider their K\"ahler potential as
\begin{equation}
K=M_{12}\left(\phi_1\phi_2 + {\rm h.c.} \right) + \cdots \;,
\end{equation}
where
\begin{equation}
 M_{12}= M\bar{\theta}^2 + M^*\theta^2 + mm^* \theta^2\bar{\theta}^2\;.
\end{equation}
Here $M\simeq M_{\rm Pl}$ and $m \simeq m_{3/2}$. Notice that the term $\phi_1\phi_2$ is allowed by the $Z_{10}$ Froggatt-Nielsen symmetry. 

Taking the string inspired axion decay constant~\cite{Svrcek:2006yi,Hui:2016ltb,Marsh:2015xka} $F_A\simeq 10^{16}\,{\rm GeV}$, we obtain
\begin{equation}
  m_A\simeq 3\times10^{-19} \sqrt{c}\left(\frac{1}{\pi^2}\right)^{5/2} \left(\frac{m_{3/2}}{1\,{\rm TeV}}\right)^{3/2} \,{\rm eV}\;.  
\end{equation}
This is a very remarkable result since $m_{3/2}\simeq 10$--$100\,{\rm TeV}$ explains the required mass for the Fuzzy DM $m_{\rm FDM}\simeq 10^{-20}$--$10^{-19}\, {\rm eV}$ as shown in the next section~\footnote{If we take the decay constant $F_A\simeq 10^{8}$--$10^{9}\,{\rm GeV}$ we may explain the recently reported high energy photon from GRB221009A~\cite{Lin:2022ocj}.}. And such a large gravitino mass, $m_{3/2}\simeq 10$--$100\,{\rm TeV}$, is free from the cosmological gravitino problem~\cite{Kawasaki:2008qe}.
 
We assume the supersymmetric (SUSY) standard model throughout this paper. If there are no SUSY particles at low energies the weak $SU(2)$ gauge coupling constant is given as $\alpha_2(M_{\rm Pl}) \simeq 1/48$ at the Planck scale which is much smaller than that in the SUSY standard model, and we obtain the mass for the EW axion, $m_A\simeq 10^{-36}\,{\rm eV}$~\cite{Lin:2022khg, McLerran:2012mm} that is too small to identify the axion with the Fuzzy DM.

\section{The Fuzzy Dark Matter}
\label{sec:fuzzy}

The Fuzzy DM of mass $10^{-20}$--$10^{-19}\,{\rm eV}$~\cite{Irsic:2017yje,Armengaud:2017nkf} is very attractive, since we may naively understand the size of galaxies by its de Broglie wavelength. Furthermore, it may not have small-scale problems including the cusp-core problem. Interestingly, the required initial value of the Fuzzy DM field to explain the DM density by its coherent oscillation is close to the string inspired decay constant $F_A\simeq 10^{16}\,{\rm GeV}$ for the axion as shown below. Such a coincidence~\cite{Hui:2016ltb} inspired many people to investigate the dynamics and astrophysical consequences of the Fuzzy DM. However, the origin of the mass was not known. We consider in this paper that the Fuzzy axion DM mass is provided by the weak $SU(2)$ instantons. And in fact, we show that in the previous section the correct mass $10^{-20}$--$10^{19}\,{\rm eV}$ is generated by the weak $SU(2)$ instantons with the decay constant $F_A\simeq 10^{16}\,{\rm GeV}$.

The abundance of Fuzzy DM is obtained from the misalignment mechanism, by considering that such ultralight boson form a condensate state $A(t)$ in the background, whose cosmic evolution follows the equation,
\begin{equation}
\label{eq:A_eom}
\ddot{A} + 3H\dot{A} + \frac{\partial V_A}{\partial A} = 0 \;,
\end{equation}
where $H(t)$ is the Hubble constant of the universe and $V_A$ is the potential in Eq.~\eqref{eq:axio-potential}. Here, notice that the leading contribution from the potential is the mass term, $\partial V_A/\partial A \simeq m_A^2 A$. When $H\gtrsim m_A$, the axion state $A(t)$ is frozen and behaves like dark energy. After $H\lesssim m_A$, the axion $A$ starts to oscillate around its VEV with an exponentially decaying amplitude.  Then, its abundance could be calculated as~\cite{Marsh:2015xka}
\begin{equation}
\Omega_A \simeq 0.08 \left(\frac{m_A}{10^{-20}\,{\rm eV}} \right)^{1/2} \left(\frac{F_A\theta_0}{10^{16}\,{\rm GeV}} \right)^2 \;,
\end{equation}
where $\theta_0 = A_0/F_A$ labels the initial misalignment angle and $A_0$ is the initial field value. If one takes $\theta_0\simeq 1$, the DM density $\Omega_{\rm CDM}h^2\simeq 0.12$~\cite{Planck:2018vyg} could be fitted by $\Omega_A$ for $m_A\simeq 10^{-19}\,{\rm eV}$ and $F_A\simeq 10^{16}\,{\rm GeV}$.

From the coupling~\eqref{eq:AWW-coupling}, one would obtain a Chern-Simons(CS)-type coupling of $A$ with electromagnetic(EM) field after the EW symmetry breaking,
\begin{equation}
\label{eq:em_cs}
\mathcal{L} \supset \frac{e^2}{32\pi^2}\frac{A}{F_A} F_{\mu\nu} \tilde{F}^{\mu\nu}\;,
\end{equation}
where $F_{\mu\nu}$ is the electromagnetic field tensor and $\tilde{F}^{\mu\nu}$ is its dual. $e$ is the $U(1)_{\rm EM}$ gauge coupling constant defined as $e= g_2 \sin \theta_{\rm W}$ and $\theta_{\rm W}$ is the EW mixing angle~\footnote{The coupling constant in Eq.~\eqref{eq:em_cs} may have an $\mathcal{O}(1)$ modification  since there could be CS-type coupling between the axion $A$ and gauge field tensors of hyper-charge $U(1)_{\rm Y}$.}.

Note that the existence of the intergalactic magnetic field may have backreactions to the Fuzzy DM due to Eq.~\eqref{eq:em_cs}. The cosmic axion field will produce magnetic helicity during cosmic evolution, which shall in turn affect the axion dynamics. As shown in Ref.~\cite{Campanelli:2005ye}, the magnetic helicity acts like a dissipation channel during the cosmic evolution of Fuzzy DM. This is described by letting $3H\to 3H+ \varepsilon$ in Eq.~\eqref{eq:A_eom}, where $\varepsilon/H \sim 10^{-26}b^2 T/{\rm GeV}$ for $F_A\simeq 10^{16}\,{\rm GeV}$ and $b$~\cite{Campanelli:2005ye} is the parameter describing the initial magnetic field. For high temperatures, $A$ behaves like dark energy in the early stage, the backreaction from the magnetic field does not change the evolution of the axion. When temperature drop below $1\,{\rm GeV}$, even for a strong magnetic field, $b\sim \mathcal{O}(1)$, the ratio $\varepsilon/H$ is extremely small, which means that the $\varepsilon$ is just a small correction to Hubble friction during the cosmic evolution of $A(t)$.  This indicates that Eq.~\eqref{eq:A_eom} is valid for the axion's cosmological evolution even in the presence of strong magnetic fields.

\section{Discussion and Conclusions}\label{sec:dis}

In this paper, we have shown that the EW axion acquires the mass of the order $10^{-20}$--$10^{-19} \,{\rm eV}$ through the electroweak instantons with the string inspired decay constant $F_A\simeq 10^{16} \,{\rm GeV}$. Therefore, it is natural to consider that the EW axion is nothing but the Fuzzy DM. The correct DM density is also obtained with the same decay constant $F_A\simeq 10^{16}\,{\rm GeV}$ for the EW axion as shown in the section~\ref{sec:fuzzy}.

 Eq.~\eqref{eq:em_cs} is a prediction of our model, which encourages observational searches for the coupling. The CS-type coupling would induce phase rotations for two helicity modes of EM waves, which would lead to a rotation on polarization angle for a linear polarized EM wave propagating through the Fuzzy DM. Different from isotropic cosmic birefringence in the CMB, this birefringence effect is local, which could be probed in the recently proposed Pulsar Polarization Arrays~\cite{Liu:2021zlt}. 

If the $B+L$ symmetry is exact, the rotation of the $B+L$ shifts the EW axion field, and hence the axion is massless as pointed out in section~\ref{sec:EW_axion}. Thus, if the EW axion is indeed the Fuzzy DM, the $B+L$ must be broken and the proton must decay through the dimension five operators \cite{Sakai:1981pk, Weinberg:1981wj}. This is also a prediction of our model. As investigated in Ref.~\cite{Evans:2021hyx}, the proton decays are predicted within the reach of JUNO and Hyper-Kamiokande.

As we have stressed in section~\ref{sec:FN_symmetry} the SUSY plays a crucial role in generating the EW axion mass of the order $10^{-20}\,{\rm eV}$. Since the SUSY standard model has a DM candidate, it is natural to have a mixed DM in the present scenario. However, the SUSY DM density depends on details of the SUSY breaking model, and hence we should take the ratio of the each DM density to be a free parameter, which can be investigated in astrophysical analysis and/or cosmology.  In this case, the EW axion should contribute partially to the DM density, from which one could obtain a bound in parameter space by letting $\Omega_A<\Omega_{\rm DM}$.

As long as $m_{3/2} \geq 10$--$100\,{\rm GeV}$ we have no gravitino problem~\cite{Kawasaki:2008qe} and as long as the axino and the scalar partner of the axion (called saxion) are heavier than the lightest SUSY particle, they decay before the BBN. However, we need special care about the saxion, since its coherent oscillation may dominate the energy density of the early universe if the initial value of the saxion is $\mathcal{O}(F_A)$, and its decay produces a large entropy in the late time. This late-time  entropy production dilutes the preexisting baryon number in the universe. However, this problem can be easily solved by imposing a ``parity" as the axion chiral multiplet ${\bar A} \to - {\bar A}$ and $W_{\mu \nu}^i\widetilde{W}^{i\mu \nu} \to - W_{\mu \nu}^i \widetilde{W}^{i\mu \nu}$~\footnote{This ``parity" requires the EW vacuum angle $\theta_{2}=0$ which might be violated. However, the cosmological evolution of the saxion does not change even if  the vacuum angle $\theta_{2}$ is not vanishing.}. Then the saxion field value is naturally set vanishing during the inflation and its coherent oscillation never occurs. The other solution is given by assuming the coupling between the axion multiplet and inflaton is relatively stronger than usual Planck suppressed operators (called the adiabatic solution)~\cite{Linde:1996cx,Nakayama:2011wqa}.

We have another cosmological problem, that is, the isocurvature problem. Since the EW axion is massless during the inflation the axion has quantum fluctuations ${\delta A} \simeq H_{\rm inf}/2\pi$ which causes too much isocurvature fluctuations if the inflation scale $H_{\rm inf}$ is too large. We obtain a constraint as $H_{\rm inf} < 10^{11}\,{\rm GeV}$~\cite{Kawasaki:2015pva} from the Planck data~\cite{Planck:2015sxf}. However, this constraint is not necessarily serious, since we have many consistent inflation models satisfying this condition~\cite{Asaka:1999jb,Nakayama:2012dw,Buchmuller:2014epa}. However, if the tensor mode $r$ in CMB is observed as $r> 10^{-7}$, our model will be excluded~\cite{Harigaya:2015yla,CMBPolStudyTeam:2008rgp}.

\begin{acknowledgements}

T. T. Y. is supported in part by the China Grant for Talent Scientific Start-Up Project and by Natural Science Foundation of China (NSFC) under grant No. 12175134 as well as by World Premier International Research Center Initiative (WPI Initiative), MEXT, Japan.

\end{acknowledgements} 

\bibliographystyle{apsrev}
\bibliography{reference}

\begin{thebibliography}{31}
\expandafter\ifx\csname natexlab\endcsname\relax\def\natexlab#1{#1}\fi
\expandafter\ifx\csname bibnamefont\endcsname\relax
  \def\bibnamefont#1{#1}\fi
\expandafter\ifx\csname bibfnamefont\endcsname\relax
  \def\bibfnamefont#1{#1}\fi
\expandafter\ifx\csname citenamefont\endcsname\relax
  \def\citenamefont#1{#1}\fi
\expandafter\ifx\csname url\endcsname\relax
  \def\url#1{\texttt{#1}}\fi
\expandafter\ifx\csname urlprefix\endcsname\relax\def\urlprefix{URL }\fi
\providecommand{\bibinfo}[2]{#2}
\providecommand{\eprint}[2][]{\url{#2}}

\bibitem[{\citenamefont{Svrcek and Witten}(2006)}]{Svrcek:2006yi}
\bibinfo{author}{\bibfnamefont{P.}~\bibnamefont{Svrcek}} \bibnamefont{and}
  \bibinfo{author}{\bibfnamefont{E.}~\bibnamefont{Witten}},
  \bibinfo{journal}{JHEP} \textbf{\bibinfo{volume}{06}}, \bibinfo{pages}{051}
  (\bibinfo{year}{2006}), \eprint{hep-th/0605206}.

\bibitem[{\citenamefont{Nomura et~al.}(2000)\citenamefont{Nomura, Watari, and
  Yanagida}}]{Nomura:2000yk}
\bibinfo{author}{\bibfnamefont{Y.}~\bibnamefont{Nomura}},
  \bibinfo{author}{\bibfnamefont{T.}~\bibnamefont{Watari}}, \bibnamefont{and}
  \bibinfo{author}{\bibfnamefont{T.}~\bibnamefont{Yanagida}},
  \bibinfo{journal}{Phys. Lett. B} \textbf{\bibinfo{volume}{484}},
  \bibinfo{pages}{103} (\bibinfo{year}{2000}), \eprint{hep-ph/0004182}.

\bibitem[{\citenamefont{Hu et~al.}(2000)\citenamefont{Hu, Barkana, and
  Gruzinov}}]{Hu:2000ke}
\bibinfo{author}{\bibfnamefont{W.}~\bibnamefont{Hu}},
  \bibinfo{author}{\bibfnamefont{R.}~\bibnamefont{Barkana}}, \bibnamefont{and}
  \bibinfo{author}{\bibfnamefont{A.}~\bibnamefont{Gruzinov}},
  \bibinfo{journal}{Phys. Rev. Lett.} \textbf{\bibinfo{volume}{85}},
  \bibinfo{pages}{1158} (\bibinfo{year}{2000}), \eprint{astro-ph/0003365}.

\bibitem[{\citenamefont{Choi et~al.}(2021)\citenamefont{Choi, Lin, Visinelli,
  and Yanagida}}]{Choi:2021aze}
\bibinfo{author}{\bibfnamefont{G.}~\bibnamefont{Choi}},
  \bibinfo{author}{\bibfnamefont{W.}~\bibnamefont{Lin}},
  \bibinfo{author}{\bibfnamefont{L.}~\bibnamefont{Visinelli}},
  \bibnamefont{and} \bibinfo{author}{\bibfnamefont{T.~T.}
  \bibnamefont{Yanagida}}, \bibinfo{journal}{Phys. Rev. D}
  \textbf{\bibinfo{volume}{104}}, \bibinfo{pages}{L101302}
  (\bibinfo{year}{2021}), \eprint{2106.12602}.

\bibitem[{\citenamefont{Lin and Yanagida}(2022{\natexlab{a}})}]{Lin:2022niw}
\bibinfo{author}{\bibfnamefont{W.}~\bibnamefont{Lin}} \bibnamefont{and}
  \bibinfo{author}{\bibfnamefont{T.~T.} \bibnamefont{Yanagida}}
  (\bibinfo{year}{2022}{\natexlab{a}}), \eprint{2208.06843}.

\bibitem[{\citenamefont{Sakai and Yanagida}(1982)}]{Sakai:1981pk}
\bibinfo{author}{\bibfnamefont{N.}~\bibnamefont{Sakai}} \bibnamefont{and}
  \bibinfo{author}{\bibfnamefont{T.}~\bibnamefont{Yanagida}},
  \bibinfo{journal}{Nucl. Phys. B} \textbf{\bibinfo{volume}{197}},
  \bibinfo{pages}{533} (\bibinfo{year}{1982}).

\bibitem[{\citenamefont{Weinberg}(1982)}]{Weinberg:1981wj}
\bibinfo{author}{\bibfnamefont{S.}~\bibnamefont{Weinberg}},
  \bibinfo{journal}{Phys. Rev. D} \textbf{\bibinfo{volume}{26}},
  \bibinfo{pages}{287} (\bibinfo{year}{1982}).

\bibitem[{\citenamefont{Froggatt and Nielsen}(1979)}]{Froggatt:1978nt}
\bibinfo{author}{\bibfnamefont{C.~D.} \bibnamefont{Froggatt}} \bibnamefont{and}
  \bibinfo{author}{\bibfnamefont{H.~B.} \bibnamefont{Nielsen}},
  \bibinfo{journal}{Nucl. Phys. B} \textbf{\bibinfo{volume}{147}},
  \bibinfo{pages}{277} (\bibinfo{year}{1979}).

\bibitem[{\citenamefont{Buchmuller and Yanagida}(1999)}]{Buchmuller:1998zf}
\bibinfo{author}{\bibfnamefont{W.}~\bibnamefont{Buchmuller}} \bibnamefont{and}
  \bibinfo{author}{\bibfnamefont{T.}~\bibnamefont{Yanagida}},
  \bibinfo{journal}{Phys. Lett. B} \textbf{\bibinfo{volume}{445}},
  \bibinfo{pages}{399} (\bibinfo{year}{1999}), \eprint{hep-ph/9810308}.

\bibitem[{\citenamefont{Choi et~al.}(2020)\citenamefont{Choi, Suzuki, and
  Yanagida}}]{Choi:2019jck}
\bibinfo{author}{\bibfnamefont{G.}~\bibnamefont{Choi}},
  \bibinfo{author}{\bibfnamefont{M.}~\bibnamefont{Suzuki}}, \bibnamefont{and}
  \bibinfo{author}{\bibfnamefont{T.~T.} \bibnamefont{Yanagida}},
  \bibinfo{journal}{Phys. Lett. B} \textbf{\bibinfo{volume}{805}},
  \bibinfo{pages}{135408} (\bibinfo{year}{2020}), \eprint{1910.00459}.

\bibitem[{\citenamefont{Hui et~al.}(2017)\citenamefont{Hui, Ostriker, Tremaine,
  and Witten}}]{Hui:2016ltb}
\bibinfo{author}{\bibfnamefont{L.}~\bibnamefont{Hui}},
  \bibinfo{author}{\bibfnamefont{J.~P.} \bibnamefont{Ostriker}},
  \bibinfo{author}{\bibfnamefont{S.}~\bibnamefont{Tremaine}}, \bibnamefont{and}
  \bibinfo{author}{\bibfnamefont{E.}~\bibnamefont{Witten}},
  \bibinfo{journal}{Phys. Rev. D} \textbf{\bibinfo{volume}{95}},
  \bibinfo{pages}{043541} (\bibinfo{year}{2017}), \eprint{1610.08297}.

\bibitem[{\citenamefont{Marsh}(2016)}]{Marsh:2015xka}
\bibinfo{author}{\bibfnamefont{D.~J.~E.} \bibnamefont{Marsh}},
  \bibinfo{journal}{Phys. Rept.} \textbf{\bibinfo{volume}{643}},
  \bibinfo{pages}{1} (\bibinfo{year}{2016}), \eprint{1510.07633}.

\bibitem[{\citenamefont{Lin and Yanagida}(2022{\natexlab{b}})}]{Lin:2022ocj}
\bibinfo{author}{\bibfnamefont{W.}~\bibnamefont{Lin}} \bibnamefont{and}
  \bibinfo{author}{\bibfnamefont{T.~T.} \bibnamefont{Yanagida}}
  (\bibinfo{year}{2022}{\natexlab{b}}), \eprint{2210.08841}.

\bibitem[{\citenamefont{Kawasaki et~al.}(2008)\citenamefont{Kawasaki, Kohri,
  Moroi, and Yotsuyanagi}}]{Kawasaki:2008qe}
\bibinfo{author}{\bibfnamefont{M.}~\bibnamefont{Kawasaki}},
  \bibinfo{author}{\bibfnamefont{K.}~\bibnamefont{Kohri}},
  \bibinfo{author}{\bibfnamefont{T.}~\bibnamefont{Moroi}}, \bibnamefont{and}
  \bibinfo{author}{\bibfnamefont{A.}~\bibnamefont{Yotsuyanagi}},
  \bibinfo{journal}{Phys. Rev. D} \textbf{\bibinfo{volume}{78}},
  \bibinfo{pages}{065011} (\bibinfo{year}{2008}), \eprint{0804.3745}.

\bibitem[{\citenamefont{Lin et~al.}(2022)\citenamefont{Lin, Yanagida, and
  Yokozaki}}]{Lin:2022khg}
\bibinfo{author}{\bibfnamefont{W.}~\bibnamefont{Lin}},
  \bibinfo{author}{\bibfnamefont{T.~T.} \bibnamefont{Yanagida}},
  \bibnamefont{and} \bibinfo{author}{\bibfnamefont{N.}~\bibnamefont{Yokozaki}}
  (\bibinfo{year}{2022}), \eprint{2209.12281}.

\bibitem[{\citenamefont{McLerran et~al.}(2012)\citenamefont{McLerran, Pisarski,
  and Skokov}}]{McLerran:2012mm}
\bibinfo{author}{\bibfnamefont{L.}~\bibnamefont{McLerran}},
  \bibinfo{author}{\bibfnamefont{R.}~\bibnamefont{Pisarski}}, \bibnamefont{and}
  \bibinfo{author}{\bibfnamefont{V.}~\bibnamefont{Skokov}},
  \bibinfo{journal}{Phys. Lett. B} \textbf{\bibinfo{volume}{713}},
  \bibinfo{pages}{301} (\bibinfo{year}{2012}), \eprint{1204.2533}.

\bibitem[{\citenamefont{Ir\v{s}i\v{c} et~al.}(2017)\citenamefont{Ir\v{s}i\v{c},
  Viel, Haehnelt, Bolton, and Becker}}]{Irsic:2017yje}
\bibinfo{author}{\bibfnamefont{V.}~\bibnamefont{Ir\v{s}i\v{c}}},
  \bibinfo{author}{\bibfnamefont{M.}~\bibnamefont{Viel}},
  \bibinfo{author}{\bibfnamefont{M.~G.} \bibnamefont{Haehnelt}},
  \bibinfo{author}{\bibfnamefont{J.~S.} \bibnamefont{Bolton}},
  \bibnamefont{and} \bibinfo{author}{\bibfnamefont{G.~D.}
  \bibnamefont{Becker}}, \bibinfo{journal}{Phys. Rev. Lett.}
  \textbf{\bibinfo{volume}{119}}, \bibinfo{pages}{031302}
  (\bibinfo{year}{2017}), \eprint{1703.04683}.

\bibitem[{\citenamefont{Armengaud et~al.}(2017)\citenamefont{Armengaud,
  Palanque-Delabrouille, Y\`eche, Marsh, and Baur}}]{Armengaud:2017nkf}
\bibinfo{author}{\bibfnamefont{E.}~\bibnamefont{Armengaud}},
  \bibinfo{author}{\bibfnamefont{N.}~\bibnamefont{Palanque-Delabrouille}},
  \bibinfo{author}{\bibfnamefont{C.}~\bibnamefont{Y\`eche}},
  \bibinfo{author}{\bibfnamefont{D.~J.~E.} \bibnamefont{Marsh}},
  \bibnamefont{and} \bibinfo{author}{\bibfnamefont{J.}~\bibnamefont{Baur}},
  \bibinfo{journal}{Mon. Not. Roy. Astron. Soc.}
  \textbf{\bibinfo{volume}{471}}, \bibinfo{pages}{4606} (\bibinfo{year}{2017}),
  \eprint{1703.09126}.

\bibitem[{\citenamefont{Aghanim et~al.}(2020)}]{Planck:2018vyg}
\bibinfo{author}{\bibfnamefont{N.}~\bibnamefont{Aghanim}} \bibnamefont{et~al.}
  (\bibinfo{collaboration}{Planck}), \bibinfo{journal}{Astron. Astrophys.}
  \textbf{\bibinfo{volume}{641}}, \bibinfo{pages}{A6} (\bibinfo{year}{2020}),
  \bibinfo{note}{[Erratum: Astron.Astrophys. 652, C4 (2021)]},
  \eprint{1807.06209}.

\bibitem[{\citenamefont{Campanelli and Giannotti}(2005)}]{Campanelli:2005ye}
\bibinfo{author}{\bibfnamefont{L.}~\bibnamefont{Campanelli}} \bibnamefont{and}
  \bibinfo{author}{\bibfnamefont{M.}~\bibnamefont{Giannotti}},
  \bibinfo{journal}{Phys. Rev. D} \textbf{\bibinfo{volume}{72}},
  \bibinfo{pages}{123001} (\bibinfo{year}{2005}), \eprint{astro-ph/0508653}.

\bibitem[{\citenamefont{Liu et~al.}(2021)\citenamefont{Liu, Lou, and
  Ren}}]{Liu:2021zlt}
\bibinfo{author}{\bibfnamefont{T.}~\bibnamefont{Liu}},
  \bibinfo{author}{\bibfnamefont{X.}~\bibnamefont{Lou}}, \bibnamefont{and}
  \bibinfo{author}{\bibfnamefont{J.}~\bibnamefont{Ren}} (\bibinfo{year}{2021}),
  \eprint{2111.10615}.

\bibitem[{\citenamefont{Evans and Yanagida}(2022)}]{Evans:2021hyx}
\bibinfo{author}{\bibfnamefont{J.~L.} \bibnamefont{Evans}} \bibnamefont{and}
  \bibinfo{author}{\bibfnamefont{T.~T.} \bibnamefont{Yanagida}},
  \bibinfo{journal}{Phys. Lett. B} \textbf{\bibinfo{volume}{833}},
  \bibinfo{pages}{137359} (\bibinfo{year}{2022}), \eprint{2109.12505}.

\bibitem[{\citenamefont{Linde}(1996)}]{Linde:1996cx}
\bibinfo{author}{\bibfnamefont{A.~D.} \bibnamefont{Linde}},
  \bibinfo{journal}{Phys. Rev. D} \textbf{\bibinfo{volume}{53}},
  \bibinfo{pages}{R4129} (\bibinfo{year}{1996}), \eprint{hep-th/9601083}.

\bibitem[{\citenamefont{Nakayama et~al.}(2011)\citenamefont{Nakayama,
  Takahashi, and Yanagida}}]{Nakayama:2011wqa}
\bibinfo{author}{\bibfnamefont{K.}~\bibnamefont{Nakayama}},
  \bibinfo{author}{\bibfnamefont{F.}~\bibnamefont{Takahashi}},
  \bibnamefont{and} \bibinfo{author}{\bibfnamefont{T.~T.}
  \bibnamefont{Yanagida}}, \bibinfo{journal}{Phys. Rev. D}
  \textbf{\bibinfo{volume}{84}}, \bibinfo{pages}{123523}
  (\bibinfo{year}{2011}), \eprint{1109.2073}.

\bibitem[{\citenamefont{Kawasaki et~al.}(2016)\citenamefont{Kawasaki, Yanagida,
  and Yokozaki}}]{Kawasaki:2015pva}
\bibinfo{author}{\bibfnamefont{M.}~\bibnamefont{Kawasaki}},
  \bibinfo{author}{\bibfnamefont{T.~T.} \bibnamefont{Yanagida}},
  \bibnamefont{and} \bibinfo{author}{\bibfnamefont{N.}~\bibnamefont{Yokozaki}},
  \bibinfo{journal}{Phys. Lett. B} \textbf{\bibinfo{volume}{753}},
  \bibinfo{pages}{389} (\bibinfo{year}{2016}), \eprint{1510.04171}.

\bibitem[{\citenamefont{Ade et~al.}(2016)}]{Planck:2015sxf}
\bibinfo{author}{\bibfnamefont{P.~A.~R.} \bibnamefont{Ade}}
  \bibnamefont{et~al.} (\bibinfo{collaboration}{Planck}),
  \bibinfo{journal}{Astron. Astrophys.} \textbf{\bibinfo{volume}{594}},
  \bibinfo{pages}{A20} (\bibinfo{year}{2016}), \eprint{1502.02114}.

\bibitem[{\citenamefont{Asaka et~al.}(2000)\citenamefont{Asaka, Hamaguchi,
  Kawasaki, and Yanagida}}]{Asaka:1999jb}
\bibinfo{author}{\bibfnamefont{T.}~\bibnamefont{Asaka}},
  \bibinfo{author}{\bibfnamefont{K.}~\bibnamefont{Hamaguchi}},
  \bibinfo{author}{\bibfnamefont{M.}~\bibnamefont{Kawasaki}}, \bibnamefont{and}
  \bibinfo{author}{\bibfnamefont{T.}~\bibnamefont{Yanagida}},
  \bibinfo{journal}{Phys. Rev. D} \textbf{\bibinfo{volume}{61}},
  \bibinfo{pages}{083512} (\bibinfo{year}{2000}), \eprint{hep-ph/9907559}.

\bibitem[{\citenamefont{Nakayama and Takahashi}(2012)}]{Nakayama:2012dw}
\bibinfo{author}{\bibfnamefont{K.}~\bibnamefont{Nakayama}} \bibnamefont{and}
  \bibinfo{author}{\bibfnamefont{F.}~\bibnamefont{Takahashi}},
  \bibinfo{journal}{JCAP} \textbf{\bibinfo{volume}{05}}, \bibinfo{pages}{035}
  (\bibinfo{year}{2012}), \eprint{1203.0323}.

\bibitem[{\citenamefont{Buchm\"uller et~al.}(2014)\citenamefont{Buchm\"uller,
  Domcke, Kamada, and Schmitz}}]{Buchmuller:2014epa}
\bibinfo{author}{\bibfnamefont{W.}~\bibnamefont{Buchm\"uller}},
  \bibinfo{author}{\bibfnamefont{V.}~\bibnamefont{Domcke}},
  \bibinfo{author}{\bibfnamefont{K.}~\bibnamefont{Kamada}}, \bibnamefont{and}
  \bibinfo{author}{\bibfnamefont{K.}~\bibnamefont{Schmitz}},
  \bibinfo{journal}{JCAP} \textbf{\bibinfo{volume}{07}}, \bibinfo{pages}{054}
  (\bibinfo{year}{2014}), \eprint{1404.1832}.

\bibitem[{\citenamefont{Harigaya et~al.}(2015)\citenamefont{Harigaya, Ibe,
  Schmitz, and Yanagida}}]{Harigaya:2015yla}
\bibinfo{author}{\bibfnamefont{K.}~\bibnamefont{Harigaya}},
  \bibinfo{author}{\bibfnamefont{M.}~\bibnamefont{Ibe}},
  \bibinfo{author}{\bibfnamefont{K.}~\bibnamefont{Schmitz}}, \bibnamefont{and}
  \bibinfo{author}{\bibfnamefont{T.~T.} \bibnamefont{Yanagida}},
  \bibinfo{journal}{Phys. Lett. B} \textbf{\bibinfo{volume}{749}},
  \bibinfo{pages}{298} (\bibinfo{year}{2015}), \eprint{1506.00426}.

\bibitem[{\citenamefont{Baumann et~al.}(2009)}]{CMBPolStudyTeam:2008rgp}
\bibinfo{author}{\bibfnamefont{D.}~\bibnamefont{Baumann}} \bibnamefont{et~al.}
  (\bibinfo{collaboration}{CMBPol Study Team}), \bibinfo{journal}{AIP Conf.
  Proc.} \textbf{\bibinfo{volume}{1141}}, \bibinfo{pages}{10}
  (\bibinfo{year}{2009}), \eprint{0811.3919}.

\end{thebibliography}

\end{document}